\documentclass[fleqn,usenatbib]{mnras}

\usepackage[T1]{fontenc}

\DeclareRobustCommand{\VAN}[3]{#2}
\let\VANthebibliography\thebibliography
\def\thebibliography{\DeclareRobustCommand{\VAN}[3]{##3}\VANthebibliography}

\usepackage{graphicx}	
\usepackage{amsmath}	
\usepackage{amssymb}	
\usepackage{newtxtext, newtxmath}

\newcommand{\RN}[1]{%
  \textup{\uppercase\expandafter{\romannumeral#1}}%
}

\title[Estimation of full-sky CMB $\hat{C_\ell}$]{An Unbiased Estimator of the Full-sky CMB Angular Power Spectrum at Large Scales using Neural Networks}

\author[P. Chanda and R. Saha]{
Pallav Chanda\thanks{E-mail: chandapallav2424@gmail.com}
and Rajib Saha\thanks{E-mail: rajib@iiserb.ac.in}
\\
Department of Physics, Indian Institute of Science Education and Research Bhopal, Bhopal-462066, Madhya Pradesh, India.
}

\date{Accepted XXX. Received YYY; in original form ZZZ}

\pubyear{2021}

\begin{document}
\label{firstpage}
\pagerange{\pageref{firstpage}--\pageref{lastpage}}
\maketitle

\begin{abstract}
Accurate estimation of the Cosmic Microwave Background (CMB) angular power spectrum is enticing due to the prospect for precision cosmology it presents.
Galactic foreground emissions, however, contaminate the CMB signal and need to be subtracted reliably in order to lessen systematic errors on the CMB temperature estimates.
Typically bright foregrounds in a region lead to further uncertainty in temperature estimates in the area even after some foreground removal technique is performed and hence determining the underlying full-sky angular power spectrum poses a challenge.
We explore the feasibility of utilizing artificial neural networks to predict the angular power spectrum of the full sky CMB temperature maps from the observed angular power spectrum of the partial sky in which CMB temperatures in some bright foreground regions are masked.
We present our analysis at large angular scales with two different masks.
We produce unbiased predictions of the full-sky angular power spectrum and recover the underlying theoretical power spectrum using neural networks.
Our predictions are also uncorrelated to a large extent.
We further show that the multipole-space covariances of the predictions of full-sky spectra made by the ANNs are much smaller than those of the estimates obtained using the pseudo-$C_\ell$ method.
\end{abstract}




\section{Introduction}
The Cosmic Microwave Background (CMB) is considered as an important probe of the early universe.
Details regarding the cosmological parameters, obtained from the angular power spectrum of the CMB temperature and polarization anisotropies, have proved useful in understanding the mechanism for formation and growth of the large-scale structure.
Satellite-based experiments like COBE \citep{bennett1996four}, WMAP \citep{bennett2003first}, and PLANCK \citep{ade2014planck}, as well as ground-based experiments like ACT \citep{sievers2013atacama}, and SPT \citep{hou2014constraints} have helped in obtaining reasonable constraints on the cosmological parameters.
Upcoming projects like the CCAT-prime \citep{stacey2018ccat}, ESA CORE \citep{de_Bernardis_2018}, and others with improved sensitivities and specialized equipment will surely make remarkable improvements in CMB measurements.

The CMB, discovered by \citet{penzias1965measurement}, is a nearly-uniform and isotropic radiation field exhibiting a virtually perfect black-body spectrum at a temperature of $2.7255 \mathrm{K}$ \citep{fixsen1996cosmic} with minute fluctuations, called temperature anisotropies, of the order of $\mu \mathrm{K}$.
During an observation, the temperature fluctuations are seen projected on the 2D surface of the spherical sky.
Therefore, the temperature anisotropies can be expressed by using a spherical harmonic expansion as follows:
\begin{equation}
    T(\theta, \phi) = \sum_{\ell=0}^{\infty} \sum_{m=-\ell}^{\ell} a_{\ell m} Y_{\ell m} (\theta, \phi),
\end{equation}
where $Y_{\ell m} (\theta, \phi)$ are the spherical harmonic functions and $a_{\ell m}$ are the harmonic modes given by,
\begin{equation}
    a_{\ell m} = \int_{\theta=0}^{\pi} \int_{\phi=0}^{2\pi} T(\theta, \phi) Y_{\ell m}^\ast (\theta, \phi) \mathrm{d}\Omega.
\end{equation}
Noting that one can measure only ($2\ell+1$) $m$-modes for a multipole, the angular power spectrum is computed as,
\begin{equation}
    \hat{C_\ell} = \frac{1}{2\ell+1} \sum_{m=-\ell}^{\ell} |a_{\ell m}|^2.
\end{equation}
$\hat{C_\ell}$ is $\chi^2$-distributed with a mean of $C_\ell^{\mathrm{th}}$, a variance of $2(C_\ell^{\mathrm{th}})^2/(2\ell+1)$, and $2\ell+1$ degrees of freedom, where $C_\ell^{\mathrm{th}}$ represents the theoretical power spectrum.
Therefore,
\begin{equation} \label{eq:fs_cl}
    C_\ell^{\mathrm{th}} = \langle \hat{C_\ell} \rangle = \frac{1}{2\ell+1} \sum_{m=-\ell}^{\ell} \langle |a_{\ell m}|^2 \rangle,
\end{equation}
where $\langle \circ \rangle$ denotes an ensemble average.
Since there is only one real sky, a realization of angular power spectrum drawn from the above-stated distribution will describe the observation of the full-sky CMB temperature field.

The statistical information present in a CMB temperature anisotropy map can be encapsulated in its angular power spectrum, $\hat{C_\ell}$, in the widely accepted structure formation model of inflation-introduced curvature perturbations that are Gaussian-distributed.
Extracting cosmological information from CMB observations is possible only when all non-cosmological signals are subtracted reliably.
Recent CMB experiments are equipped with adequate angular resolution and sensitivity to probe CMB anisotropies at large angular scales.
Therefore, the primary source of uncertainty in estimates is due to the galactic foreground emissions.

The galactic foregrounds are characterized by higher intensities (temperatures) in regions around the galactic plane compared with other regions of the sky.
This results in comparatively higher uncertainties in CMB measurements in those regions when reconstructed using existing methods like COMMANDER \citep{2004ApJS..155..227E, 2008ApJ...676...10E}, ILC \citep{Saha_2006, PhysRevD.78.023003, Sudevan_2018}, etc.
Bright sources in some sky regions also result in further unreliability of the measurements.
Applying a mask on the CMB temperature map, that excludes the components in the `bright' foreground regions, aids data analysis.
Angular power spectrum of the partial sky then summarizes the information present in the masked sky map.
However, recovering an unbiased estimator of the underlying full-sky angular power spectrum is an important problem in cosmology to solve.

Common maximum likelihood methods \citep{PhysRevD.57.2117, PhysRevD.67.023001}, Gibbs and Bayesian sampling methods \citep{2004ApJS..155..227E, 10.1093/mnras/stv2501} exist for estimating the full-sky CMB temperature anisotropy angular power spectrum, $\hat{C_\ell}$, from angular power spectrum of the finite-area cut-sky map, $\hat{\Tilde{C_\ell}}$.
However, these methods involve complex computations and are CPU expensive since they scale as $\ell_\mathrm{max}^6$, with $\ell_\mathrm{max}$ being the maximum multipole.
The method using Gabor transforms, introduced by \citet{hansen2002gabor}, is again hindered by slow calculations of the correlation matrix of $\hat{\Tilde{C_\ell}}$ needed for their maximum likelihood analysis.

Furthermore, \citet{1973ApJ...185..413P}, \citet{PhysRevD.64.083003} and \citet{Hivon_2002} put forward the pseudo-$C_\ell$ algorithm for estimation of the full-sky angular power spectrum from the masked-sky spectrum, which foregoes slow calculations by introducing a mode-mode coupling kernel that depends only on the geometry of the mask used.
Various extensions of the pseudo-$C_\ell$ method exist like those presented by \citet{refId0}, \citet{10.1093/mnras/stw2752}, and the references therein.
Nevertheless, the $\hat{C_\ell}$ estimates using these methods have large error-bars on lower multipoles even after using substantial bin-width, and are also limited in the sense that the unbiased estimator relies on a linear transformational relation that exists between the ensemble averages, $\langle \hat{\Tilde{C_\ell}} \rangle$ and $\langle \hat{C_\ell} \rangle$, which may not necessarily be the best possible relation between $\hat{\Tilde{C_\ell}}$ and $\hat{C_\ell}$ of the individual realizations.
Therefore, the requirement for a technique that is efficient, reliable, and enables optimal recovery of the full-sky angular power spectrum at all multipoles is well documented.

In this work, we have explored the use of Artificial Neural Networks (ANNs) to predict the full-sky CMB angular power spectrum based on the power spectrum of the partial (or masked) sky as a new alternative method.
We are interested in reconstruction of the large-scale CMB temperature anisotropy power spectrum in this initial article.
Using random realizations of the CMB power spectrum, we simulate maps of CMB temperature anisotropy.
Choosing suitable temperature masks to apply on these maps, we calculate angular power spectra of the masked maps as well as the unmasked (full-sky) maps, hereafter represented as $\hat{\Tilde{C_\ell}}$s and $\hat{C_\ell}$s respectively.
The simulated data is used to construct training and testing sets for the ANNs.
The training data plays a role similar to that of the priors, and the neural network uses it to learn the complicated mapping that exists between the $\hat{\Tilde{C_\ell}}$ and $\hat{C_\ell}$ of individual realizations.
ANNs are well-known as universal function approximators \citep[see][]{HORNIK1991251, pinkus_1999}.
Using this novel method enables us to get unbiased predictions of the full-sky spectra without binning at lower multipoles and with a minimal bin width at higher multipoles.
We also obtain significantly smaller error-bars on our full-sky $\hat{C_\ell}$ predictions.
The ANN predictions help acquire an unbiased estimator of the underlying theoretical $C_\ell^{\mathrm{th}}$ as well.

This paper is organized as follows:
In Section \ref{sec:estimator}, we derive the necessary equations that relate the full-sky angular power spectrum to the partial-sky power spectrum.
Next, we give a brief review of the concept of Artificial Neural Networks focused on the context of our application in Section \ref{sec:ANN}.
In Section \ref{sec:methodology}, we present our strategy for getting an unbiased estimate of the full-sky angular power spectrum from power spectrum of the partial-sky.
Section \ref{sec:simulations} describes the process for simulating CMB maps, and the corresponding $\hat{C_\ell}$s and $\hat{\Tilde{C_\ell}}$s.
We list our methodology, the procedure for training neural networks, and the binning strategy required to get an unbiased estimate after making predictions in Section \ref{sec:training}.
Section \ref{sec:results} presents the results of our analyses on simulated data.
Our conclusions and possible future work on our method are discussed in Section \ref{sec:conclusions}.

\section{From Partial-sky to the Full-sky Power Spectrum} \label{sec:estimator}
The effects of a mask on the full-sky temperature field can be described by a position-dependent weighting using a window function $W$.
The effect of a finite window function on the temperature field is given by \citep{PhysRevD.64.083003},
\begin{equation}
    \Tilde{T}(\theta, \phi) = W(\theta, \phi) T(\theta, \phi).
\end{equation}
Defining the harmonic space window function, $W_{\ell m}^{\ell' m'}$, as:
\begin{equation}
    W_{\ell m}^{\ell' m'} = \int_{\theta=0}^{\pi} \int_{\phi=0}^{2\pi} Y_{\ell' m'} (\theta, \phi) W(\theta, \phi) Y_{\ell m}^\ast (\theta, \phi) \mathrm{d}\Omega,
\end{equation}
the harmonic modes of the partial (masked) sky are given by,
\begin{equation} \label{eq:ps_harmonic}
    \Tilde{a}_{\ell m} = \sum_{\ell' m'} W_{\ell m}^{\ell' m'} a_{\ell' m'}.
\end{equation}

Using these harmonic modes, the angular power spectrum of the masked sky can be calculated as:
\begin{equation} \label{eq:ps_cl}
    \hat{\Tilde{C_\ell}} = \frac{1}{2\ell+1} \sum_{m=-\ell}^{\ell} |\Tilde{a}_{\ell m}|^2.
\end{equation}
Thus, one can relate the partial-sky power spectra to the full-sky spectra by taking ensemble averages on both sides and then using equation \ref{eq:ps_harmonic} and equation \ref{eq:fs_cl} as follows:
\begin{align}
    \langle \hat{\Tilde{C_\ell}} \rangle &= \frac{1}{2\ell+1} \sum_{m=-\ell}^{\ell} \langle \Tilde{a}_{\ell m} \Tilde{a}_{\ell m}^{\dagger} \rangle \nonumber \\
    &= \sum_{\ell'} \sum_{mm'} W_{\ell m}^{\ell' m'} \langle \hat{C_{\ell'}} \rangle (W_{\ell m}^{\ell' m'})^{\dagger}.
\end{align}
Simplifying the above expression by using a matrix $M$ in order to describe the mode-mode coupling \citep{Hivon_2002}, it becomes:
\begin{equation} \label{eq:coupling}
    \langle \hat{\Tilde{C_\ell}} \rangle = \sum_{\ell'} M_{\ell \ell'} \langle \hat{C_{\ell'}} \rangle.
\end{equation}
At the large scales that we are working on, the instrumental noise is negligible in magnitude and hence, ignoring it is a suitable assumption.
Thus, inverting equation \ref{eq:coupling}, the true full-sky power spectra can be represented in terms of the pseudo partial-sky spectra,
\begin{equation} \label{eq:coupling_inv}
    \langle \hat{C_\ell} \rangle = \sum_{\ell'} M_{\ell \ell'}^{-1} \langle \hat{\Tilde{C_{\ell'}}} \rangle.
\end{equation}

For a given realization of the CMB sky, we use the convention that $\mathbf{c}$ represents the column vector having $\hat{C_\ell}$ ($\ell = 0, 1, ..., \ell_{\mathrm{max}}$) as its elements and $\mathbf{\Tilde{c}}$ represents the column vector having $\hat{\Tilde{C_\ell}}$ ($\ell = 0, 1, ..., \ell_{\mathrm{max}}$) as its elements, where $\ell_{\mathrm{max}}$ depends on resolution of the temperature maps, throughout the paper.
If the ensemble averages in equation \ref{eq:coupling_inv} are eliminated, the resulting relation qualifies as an estimator of the full-sky CMB $\hat{C_\ell}$ using partial-sky $\hat{\Tilde{C_\ell}}$.
This relation is key to the existing pseudo-$C_\ell$ methods.
However, the linear transformational relation holds only on the stated ensemble averages and thus, inherently, excludes the complex functional relationships that may exist between the full-sky and partial-sky power spectrum of individual realizations of the CMB sky.
Thus, to get a better estimator, we also replace the linear transformation with a functional relation while discarding the ensemble averages. equation \ref{eq:coupling_inv} can, then, be generalized as follows:
\begin{equation} \label{eq:function_cls}
    \mathbf{c} = f(\mathbf{\Tilde{c}}).
\end{equation}

Describing the above equation by using an equivalent mapping notation, we obtain,
\begin{equation} \label{eq:mapping_cls}
    \mathbf{\Tilde{c}} \xrightarrow{\ f\ } \mathbf{c}.
\end{equation}
Based on this expression, we explore a strategy using ANNs, discussed in Section \ref{sec:methodology}, to approximate the full-sky CMB angular power spectrum.

\section{Artificial Neural Networks} \label{sec:ANN}
Artificial Neural Networks (ANNs) are computational information processing systems aimed at recognizing underlying relationships in a set of data.
Historically, the learning mechanisms of neural networks have drawn inspiration from those of the human brain and the nervous system.
However, the progressive study of neural networks and discovery of various architectures that had little connection with working of the brain have led to a new outlook towards neural networks in the modern age.
ANNs have found a wide range of applications ranging from financial predictions and function approximation to computer vision and speech recognition/synthesis.
In this section, we provide a brief overview of ANNs and concepts pertaining to our use-case.
The reader may consider referring to \citet{bishop1995neural} for a detailed description of neural networks.

\begin{figure}
    \centering
    \includegraphics[width=.48\textwidth]{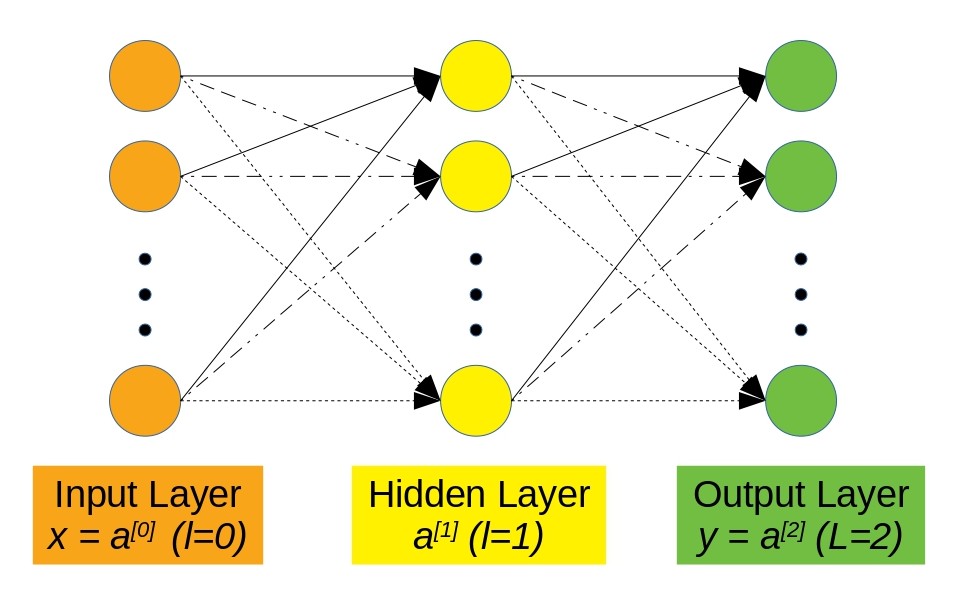}
    \caption{Figure showing a `dense' ANN with an input layer, an output layer and a single hidden layer.
    Each circular unit represents a `neuron' and the lines connecting them represent the associated weights.
    The word `dense' implies a fully-connected network in which all neurons in any layer of the network are connected to all neurons in the subsequent layer.}
    \label{fig:ANN}
\end{figure}

In our work, we focus on supervised learning with `dense' (fully-connected) feed-forward ANNs.
An ANN consists of an input layer, an output layer and one or more hidden layers (Fig. \ref{fig:ANN}).
Each of the circular units is called a `neuron' and the lines connecting different neurons represent the associated weights.
Consider the input features in a particular training example to be represented as a column vector $\mathbf{x}$ whose elements are $x_p$ ($p = 1, ..., n_{\mathrm{in}}$, where $n_{\mathrm{in}}$ is the number of input features) and the true output values (or ground truths) to be represented as a column vector $\mathbf{\hat{y}}$ whose elements are $\hat{y}_q$ ($q = 1, ..., n_{\mathrm{out}}$, where $n_{\mathrm{out}}$ is the number of output values).
An ANN can then be understood as a mapping from input to output, 

\begin{equation}
    \mathbf{x} \xrightarrow{\mathrm{\ ANN\ }} \mathbf{y}.
\end{equation}

\subsection{Forward Propagation}

The ANN learns the mapping by training the weights and biases in the network.
Each neuron in a layer is connected to all neurons of the previous layer.
We use the convention that superscript $[l]$ represents the $l$-th layer.
Thus, any layer $l$ has $n^{[l]}$ neurons, associated with $\mathbf{W}^{[l]}$ weights [$w_{ij}^{[l]}$ joining neuron $i$ in layer $l$ and neuron $j$ in layer ($l-1$)] and $\mathbf{b}^{[l]}$ biases (or offsets) [$b_{i}^{[l]}$ of neuron $i$ in layer $l$].
A neuron comprises of a linear part and an activation part.
Let the vectors $\mathbf{z}^{[l]}$ and $\mathbf{a}^{[l]}$ represent the linear and activation parts of the neurons in the $l$-th hidden layer, respectively.
Then, the \textit{forward propagation} step is given by,
\begin{align}
    z_i^{[l]} &= \sum_{j=1}^{n^{[l-1]}} w_{ij}^{[l]} a_{j}^{[l-1]} + b_{i}^{[l]},\\
    a_i^{[l]} &= g^{[l]} \left( z_i^{[l]} \right),
\end{align}
where $g^{[l]} \left( z \right)$ represents the activation function $g^{[l]}$ of the $l$-th layer.
The activation functions are chosen to be non-linear functions, e.g. logistic sigmoid, tanh, ReLU, etc., to introduce a non-linearity in the network.
By convention, the input layer is referred to as the zeroth layer, and $\mathbf{a}^{[0]} = \mathbf{x}$.
The hidden layers, along with the activation functions, help the ANN to create complex non-linear mappings.

For regression problems in which the output features are real-valued numbers, the output later has an identity activation function.
If the output layer is enumerated as $L$, the corresponding activations can be computed as:
\begin{equation}
    y_q = a_q^{[L]} = z_q^{[L]} = \sum_{j=1}^{n^{[L-1]}} w_{qj}^{[L]} a_{j}^{[L-1]} + b_{q}^{[L]}.
\end{equation}
This constitutes the network's prediction vector, $\mathbf{y}$, which depends on the weights, biases, and activation functions of all the layers in the network, thus creating a map.

\subsection{The Training Process}

Consider training an ANN with $m$ training examples of input and true output vector pairs ($\mathbf{x}^{(k)}, \mathbf{\hat{y}}^{(k)}$), where we use the convention that superscript $(k)$ represents the $k^{\mathrm{th}}$ training example.
The ANN \textit{forward propagates} to get the neuron activations for all $m$ training examples and eventually the predicted outputs $\mathbf{y}^{(k)}$ ($k = 1, ..., m$).
A training algorithm that minimizes a \textit{cost function} ($J$) trains an ANN to predict the outputs.

For machine learning to be applicable in cosmology, precise error estimation becomes an important factor to consider.
For regression problems, uncertainty representation with feed-forward neural networks can be addressed by using a Bayesian approach.
This uncertainty can be categorised as:
1. \textit{aleatoric} uncertainty arising from inherent noise in the data, and
2. \textit{epistemic} uncertainty emanating from our ignorance about the model parameters.
Aleatoric uncertainty cannot be minimised even if more data is collected, however, epistemic uncertainty can be reduced given enough data \citep{kendall2017uncertainties}.
Implementing Concrete dropout \citep{gal2017concrete} in layers of a neural network is one way to ensure a Bayesian model.
\textit{Dropout} can be understood as training many neural networks with different architectures in parallel achieved by randomly dropping neurons, and their connections from a single model.
Dropout used at training time helps prevent neural networks from overfitting as shown by \citet{srivastava2014dropout}.
\citet{gal2016dropout} have shown that dropout employed at test time can be treated as a Bayesian approximation to estimate the network's epistemic uncertainty over multiple evaluations, with the caveat being a proper tuning of the dropout probability.
\citet{gal2017concrete} address this limitation with Concrete dropout that makes the dropout probability a learnable parameter during network training ensuring its appropriate tuning.
To estimate the aleatoric uncertainty, the network is enabled to also predict the error in $\mathbf{y}$ as shown by \citet{kendall2017uncertainties}.
In practice, the model outputs the predicted error (log of variance, for numerical stability) vector $\mathbf{s}$ in addition to the prediction vector $\mathbf{y}$.
The network is, thus, trained using the \textit{loss} function,
\begin{equation}\label{eq:loss_func}
    \mathcal{L} = \frac{1}{n_\mathrm{out}} \sum_{q=1}^{n_\mathrm{out}} \left[ \frac{1}{2} \exp{(-s_q)} || y_q - \hat{y}_q ||^2 + \frac{1}{2} s_q \right],
\end{equation}
where $s_q = \log \sigma_q^2$ is the predicted log variance. Thus, one obtains the \textit{cost} function by averaging the above loss function over all training examples,
\begin{equation}
    J = \frac{1}{m} \sum_{k=1}^{m} \mathcal{L}^{(k)}.
\end{equation}

In order for the network to learn the mapping, it needs to adjust its weights and biases to minimize the \textit{cost} ($J$).
A technique called \textit{Error Back Propagation} \citep{hecht1992theory} estimates the gradients of the weights and biases with respect to the cost function based on training examples that the network is currently learning from.
Taking a small step from $w$ to $w + \delta w$ and from $b$ to $b + \delta b$ changes the cost by an amount $\delta J$.
The task of updating the parameters by a minimal amount in every iteration of training is handled by an \textit{Optimization Algorithm}.
Various types of optimization algorithms exist, for e.g. \textit{Gradient Descent}, \textit{Stochastic Gradient Descent (SGD)}, \textit{Momentum}, \textit{RMSProp} \citep{hinton2012neural}, \textit{Adaptive Moment Estimation (ADAM)} \citep{kingma2014adam}, etc., with their own sets of pros and cons.
It is beyond the scope of this paper to provide a review of these, and we refer the reader to \citet{8903465} for the same.

Updating the weights of a neural network by estimating gradients based on all training examples during each iteration of network-training may lead the neural network to learn at a slow pace or get stuck in local minima of the high-dimensional weight space.
To tackle this problem, it is common to use mini-batches of size $m_b\ (< m)$.
Fluctuations in the weight space brought about by the mini-batch optimization algorithms like \textit{Mini-batch Stochastic Gradient Descent (MSGD)}, \textit{ADAM}, etc. by computing the gradients based on a subset of the entire training set during each iteration allow the neural network to jump to another possible minimum as shown by \citet{ruder2016overview}.
In literature, the terminology `epoch' addresses the number of iterations of optimization algorithm in which the network `sees' all training examples once.
Thus,
\begin{equation}
    1 \mathrm{\ epoch} = \left\lceil \frac{m}{m_b} \right\rceil \mathrm{\ iterations,}
\end{equation}
where $m_b$ represents the number of training examples in a batch (i.e., batch-size) and $\lceil \circ \rceil$ represents the ceiling function.

The training process for our neural networks can, thus, be summed up as follows:
\begin{enumerate}
    \item Randomly initialize all the weights and initialize all the biases to zero.
    \item Shuffle all examples in the training set and split it into $n_\mathrm{bat}$ batches, each having $m_b$ examples (except the last batch, which can have less than $m_b$ examples). \label{step:train2}
    \item
    \begin{enumerate}
        \item \textit{Forward propagate} to get the neuron activations using training examples in the considered batch. \label{step:train3a}
        \item Compute the \textit{cost} on the current batch ($J_r$) as:
        \begin{equation}
            \hspace{\parindent} J_r = \sum_{k=1}^{\mathrm{m_b}} \mathcal{L}^{(k)},
        \end{equation}
        where subscript $r$ denotes the current batch.
        \item \textit{Back propagate} errors to get estimates of gradients of the weight space, $\frac{\partial J_r}{\partial w}$ and $\frac{\partial J_r}{\partial b}$.
        \item Update the weights and the biases using an optimization algorithm and the gradients. \label{step:train3d}
        \item Repeat steps \ref{step:train3a} to \ref{step:train3d} for all batches in the training set (i.e., $\left\lceil \frac{m}{m_b} \right\rceil$ iterations).
        \item Compute the total $cost$ ($J_{\mathrm{tot}}$) as:
        \begin{equation}
            \hspace{\parindent} J_{\mathrm{tot}} = \frac{1}{m} \sum_{r=1}^{n_{\mathrm{bat}}} J_r
        \end{equation}
    \end{enumerate} \label{step:train3}
    \item Repeat steps \ref{step:train2} and \ref{step:train3} for $t$ number of epochs or till $J_{\mathrm{tot}}$ is minimized.
\end{enumerate}

It is necessary to manually tune the hyper-parameters of an ANN model for training it successfully.
As a general rule of thumb, implementing Concrete dropout is achieved by having a network with a large number of neurons in each layer, and letting the network choose the number of neurons it wants to utilize based on the dropout probability that it learns during training.
Tracking the performance of a neural network on a separate validation set while training ensures that the network does not over-fit training data.
Alongside, applying dropout and regularisation methods also assist in reducing over-fitting.
The ANN, with its trained weights and biases, is then used for making predictions on the test set.
This ensures that testing happens on examples that the ANN has never seen before, resulting in a fair assessment of the network's performance.

\section{The Full-sky CMB Angular Power Spectrum Estimator} \label{sec:methodology}
In view of the discussions in Section \ref{sec:ANN}, an artificial neural network is capable of learning a mapping from input to output given some training data.
The trained network is then able to predict the outputs based on the input data.
Thus, an ANN can be employed to figure out the mapping from $\mathbf{\Tilde{c}}$ to $\mathbf{c}$, which we arrived at in equation \ref{eq:mapping_cls} of Section \ref{sec:estimator}.
Representing in the form of an expression, we write,
\begin{equation}
    \mathbf{\Tilde{c}} \xrightarrow{\mathrm{\ ANN\ }} \mathbf{c}.
\end{equation}

The ANN learns the mapping by training on simulated data using $(\mathbf{x}^{(k)}, \mathbf{\hat{y}}^{(k)}) = (\mathbf{\Tilde{c}}^{(k)}, \mathbf{c}^{(k)})$, where $k = 1, ..., m$ represent the $m$ training examples.
Partial-sky $\hat{\Tilde{C_\ell}}$ at all multipoles are required for estimating the full-sky $\hat{C_\ell}$ at each multipole, which leads us to opt for dense feed-forward ANNs.
Estimating the errors in the $\hat{C_\ell}$ predictions is also crucial.
Hence, we implement Concrete dropout in our ANN model.
The network is made to predict the error in the $\hat{C_\ell}$ prediction along with the $\hat{C_\ell}$, and trained using the loss function stated in equation \ref{eq:loss_func}.
To get a combined estimate of both types of uncertainty, we iterate the network several times with dropout for the same input features.
During each iteration:
\begin{enumerate}
    \item The standard deviation obtained from the aleatoric uncertainty estimate of the network is used to generate a normal distribution.
    \item A random noise realization is drawn from it.
    \item This noise realization is added to the $\hat{C_\ell}$ predicted by the neural network.
\end{enumerate}
The final output is then obtained by computing the mean and standard deviation of the set of $\hat{C_\ell}$ realizations from all iterations.

A rich training data set, containing several realizations of the partial-sky and full-sky spectra, is primary to training a good model for predicting the full-sky spectrum from partial-sky spectrum.
In the following sections, we discuss how we obtain the simulated data and build the network.

\section{Simulations of the Angular Power Spectra} \label{sec:simulations}
We make use of
\href{https://healpix.sourceforge.io/}{\textsc{HEALPix}}\footnote{https://healpix.sourceforge.io/} \citep{gorski2005healpix} software in python (\href{https://github.com/healpy/healpy}{\texttt{healpy}}\footnote{https://github.com/healpy/healpy}) to simulate and analyze random realizations of the CMB temperature anisotropy.
\texttt{healpy.sphtfunc.synfast} can simulate full-sky CMB maps given the theoretical $C_\ell^{\mathrm{th}}$.
We utilize the \textsc{Planck} mission's theoretical CMB temperature anisotropy power spectrum \citep{planck_cosmo_param} as the input theoretical $C_\ell^{\mathrm{th}}$ for map generation.
The cosmological parameters that generated this theoretical power spectrum model were the best-fitting parameters obtained using a standard $\Lambda$CDM model with a power-law spectral index \citep{planck_cosmo_param}:\\
$\{\Omega_b h^2, \Omega_c h^2, h, \tau, n_s, \log_e(10^{10} A_s)\}$\\
$\mbox{\qquad \qquad \qquad \qquad \quad \ \ } = \{0.022, 0.12, 0.673, 0.054, 0.966, 3.045\}$,\\
where $\Omega_b$ is the baryonic energy density, $\Omega_c$ is the dark matter density, $h$ is the Hubble constant (in units of $100$ km/s/Mpc), $n_s$ is the scalar spectral index, $\tau$ is the optical depth to the decoupling surface, and $A$ is the parameter that characterises the amplitude of the initial perturbations.

\begin{figure}
    \centering
    \includegraphics[width=.48\textwidth]{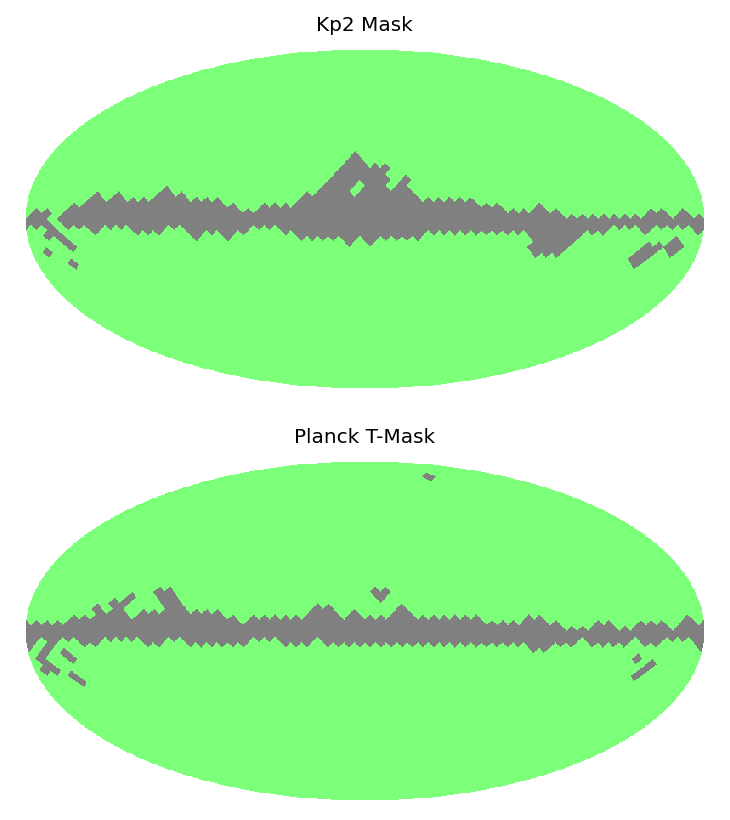}
    \caption{The Kp2 mask and the Planck T-mask at $N_{\mathrm{side}} = 16$ that we use for our analyses.
    The masked region of the sky is shown in gray colour and the unmasked region is shown in light green colour.}
    \label{fig:masks}
\end{figure}

We work with low resolution maps at \textsc{HEALPix} pixel resolution $N_{\mathrm{side}}=16$, providing an angular resolution of $\approx 219.87$ arcmin.
At this resolution, only large angular scale theoretical $C_\ell^{\mathrm{th}}$ up to $\ell_{\mathrm{max}} = 2 N_{\mathrm{side}} = 32$ are provided as input to \texttt{synfast} for map generation.
We also smooth the simulated CMB maps with a Gaussian beam having FWHM = $540$ arcmin (roughly $2.46$ times the angular resolution of the maps) to ensure a band-limited signal.
Passing the above information as arguments to \texttt{synfast} and considering the required pixel-window smoothing, we generate a (random) realization of the CMB signal.

\begin{figure*} 
    \centering
    \includegraphics[width=.99\textwidth]{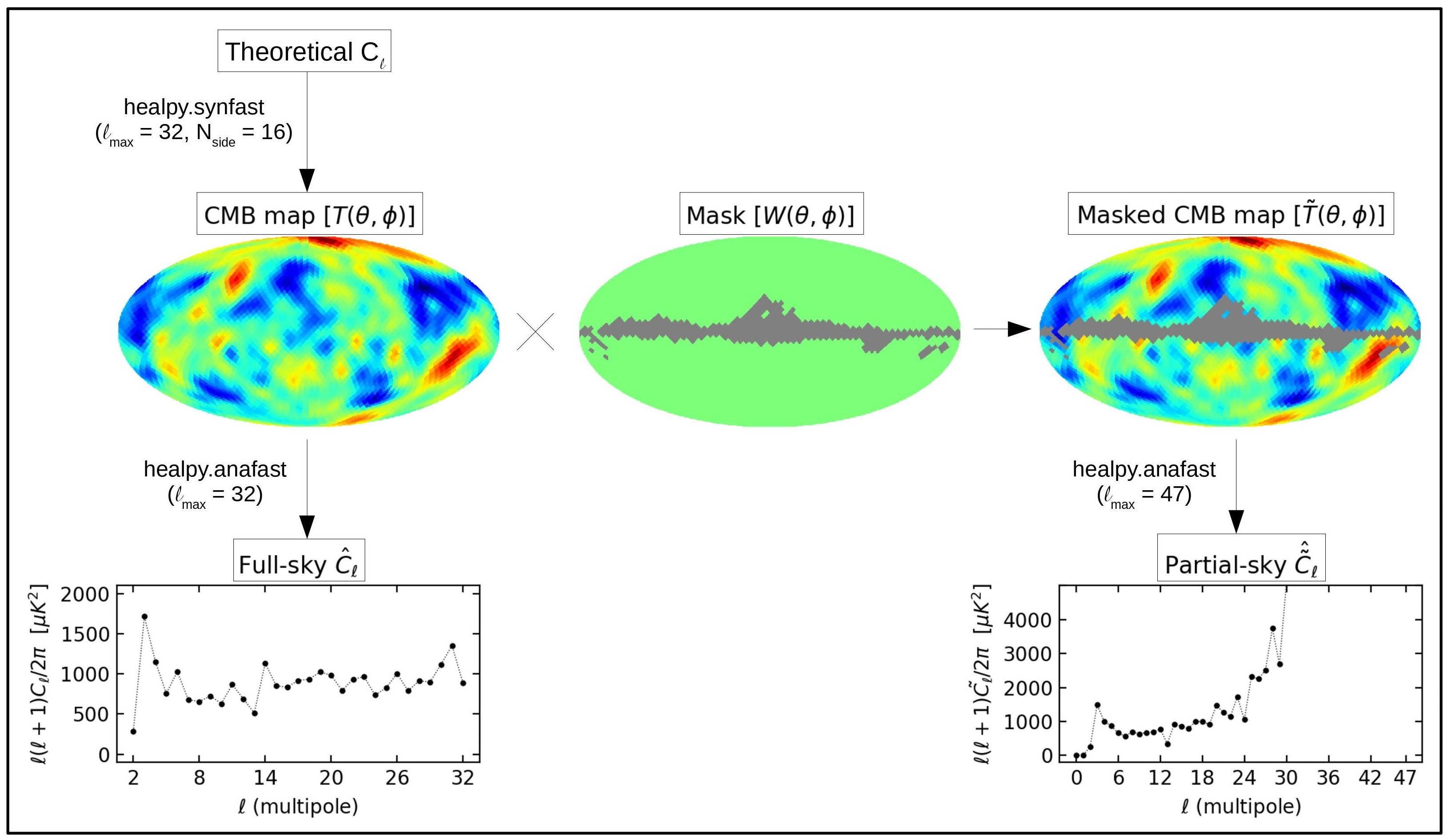}
    \caption{Figure showing the procedure for generating the simulated data at $N_{\mathrm{side}}=16$.
    A realization of the CMB sky is obtained given the theoretical power spectrum, $C_\ell^{\mathrm{th}}$.
    Its angular power spectrum gives the $\hat{C_\ell}$.
    We then apply the desired mask on the CMB map.
    Angular power spectrum of the masked CMB map gives the $\hat{\Tilde{C_\ell}}$.
    The $\hat{\Tilde{C_\ell}}$ are very large at higher multipoles due to the loss of information caused by masking.}
    \label{fig:procedure_mask}
\end{figure*}

Our analysis depicting prediction of the full-sky angular power spectrum from power spectrum of the cut-sky presents the use of two different masks to obtain the partial-sky signal maps.
We use the `\href{https://lambda.gsfc.nasa.gov/product/map/dr1/imask.cfm}{Kp2 mask}\footnote{https://lambda.gsfc.nasa.gov/product/map/dr1/imask.cfm}', with $\approx 15\%$ masked pixels, as our first mask.
It is available as a LAMBDA product at $N_{\mathrm{side}} = 512$, which we downgrade to $N_{\mathrm{side}} = 16$.
All the non-integer values that arise as a result of downgrading are rounded off to either 0 or 1.
We use the \href{https://irsa.ipac.caltech.edu/data/Planck/release\_3/all-sky-maps/previews/COM\_CMB\_IQU-commander\_2048\_R3.00\_full/index.html}{temperature mask}\footnote{https://irsa.ipac.caltech.edu/data/Planck/release\_3/all-sky-maps/previews/COM\_CMB\_IQU-commander\_2048\_R3.00\_full/index.html} given alongside Planck PR3 CMB IQU maps produced by the COMMANDER pipeline as our second mask having $\approx 12\%$ masked pixels.
Hereafter, we refer to it as the `Planck~T-mask'.
It is available at $N_{\mathrm{side}} = 2048$, and we downgrade it to $N_{\mathrm{side}} = 16$.
The mask also goes through a similar processing as the Kp2 mask.
We apply these raw ($1/0$) masks (shown in Fig. \ref{fig:masks}) to realizations of the CMB sky for getting the masked CMB maps.

The angular power spectra ($\hat{C_\ell}$s) of the simulated full-sky maps are computed using \texttt{healpy.sphtfunc.anafast}, again setting $\ell_{\mathrm{max}}=32$ as the maps were generated using the same $\ell_{\mathrm{max}}$ on $C_\ell^{\mathrm{th}}$.
However, in the case of partial-sky maps, we compute the angular power spectra ($\hat{\Tilde{C_\ell}}$s) with $\ell_{\mathrm{max}} = 3 N_{\mathrm{side}} - 1 = 47$ so as to extract as much information as possible from the masked maps, assisting the ANNs in making better predictions.
It is customary to plot the CMB $\hat{C_\ell}$ as $\ell(\ell+1)C_\ell/2\pi$ which brings out a lot of the structure at smaller scales.
Additionally, we divide the power spectra by a factor of $B_\ell^2 P_\ell^2$, where $B_\ell$ represents the beam window function of the Gaussian beam corresponding to FWHM = $540$ arcmin and $P_\ell$ represents the pixel window function for $N_{\mathrm{side}}=16$, so as to account for the effects of smoothing and pixel area, respectively.
Fig. \ref{fig:procedure_mask} depicts the procedure for obtaining the required data for one realization of the CMB sky.

Following the procedure, we simulate $1.2\times10^5$ random realizations of the CMB signal.
Getting their angular power spectra give us examples of the full-sky CMB $\hat{C_\ell}$.
Then, we mask all the full-sky maps with the Kp2 mask and compute their angular power spectra to get $1.2\times10^5$ examples of the corresponding partial-sky CMB $\hat{\Tilde{C_\ell}}$.
Further, we use the Planck T-mask on all the full-sky maps and obtain examples of the corresponding partial-sky CMB $\hat{\Tilde{C_\ell}}$.
This constitutes the simulated data that we use to train our neural networks.

\section{Training the Networks and Making Predictions} \label{sec:training}
Due to the statistical nature of the problem, it is known that a neural network needs to `see' at least $10^5$ training examples to learn the mapping of possible feature variations.
Out of the $1.2\times10^5$ $(\mathbf{\Tilde{c}}, \mathbf{c})$ pairs (examples of the partial-sky $\hat{\Tilde{C_\ell}}$ and the full-sky $\hat{C_\ell}$) obtained using the Kp2 mask, $10^5$ pairs form the training set, while $10^4$ pairs each form the validation and test sets for training the network.
A similar train, validation, and test set splitting is performed for $(\mathbf{\Tilde{c}}, \mathbf{c})$ pairs obtained using the Planck T-mask.

\begin{figure}
    \centering
    \includegraphics[width=.48\textwidth]{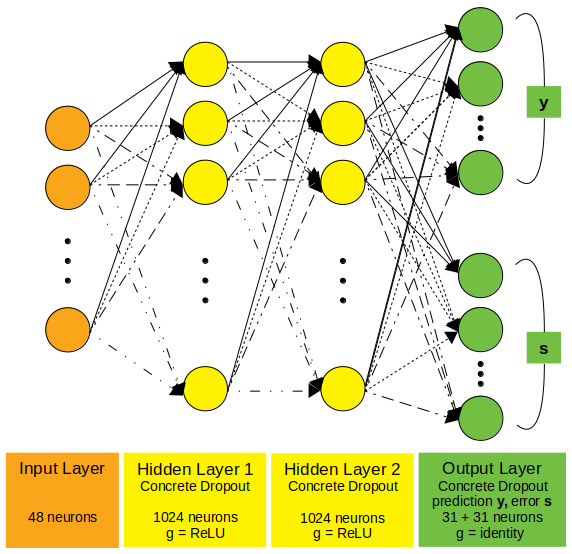}
    \caption{Representation of the ANN architecture that we use for our analyses.
    The ANN has an input layer with 48 neurons and two hidden layers with 1024 neurons each and ReLU activation function.
    The output layer comprises of a prediction vector $\mathbf{y}$ and an error prediction vector $\mathbf{s}$ with 31 neurons each.
    The hidden layers and the output layer are implemented with Concrete dropout.}
    \label{fig:our_ann}
\end{figure}

\begin{figure}
    \centering
    \includegraphics[width=.48\textwidth]{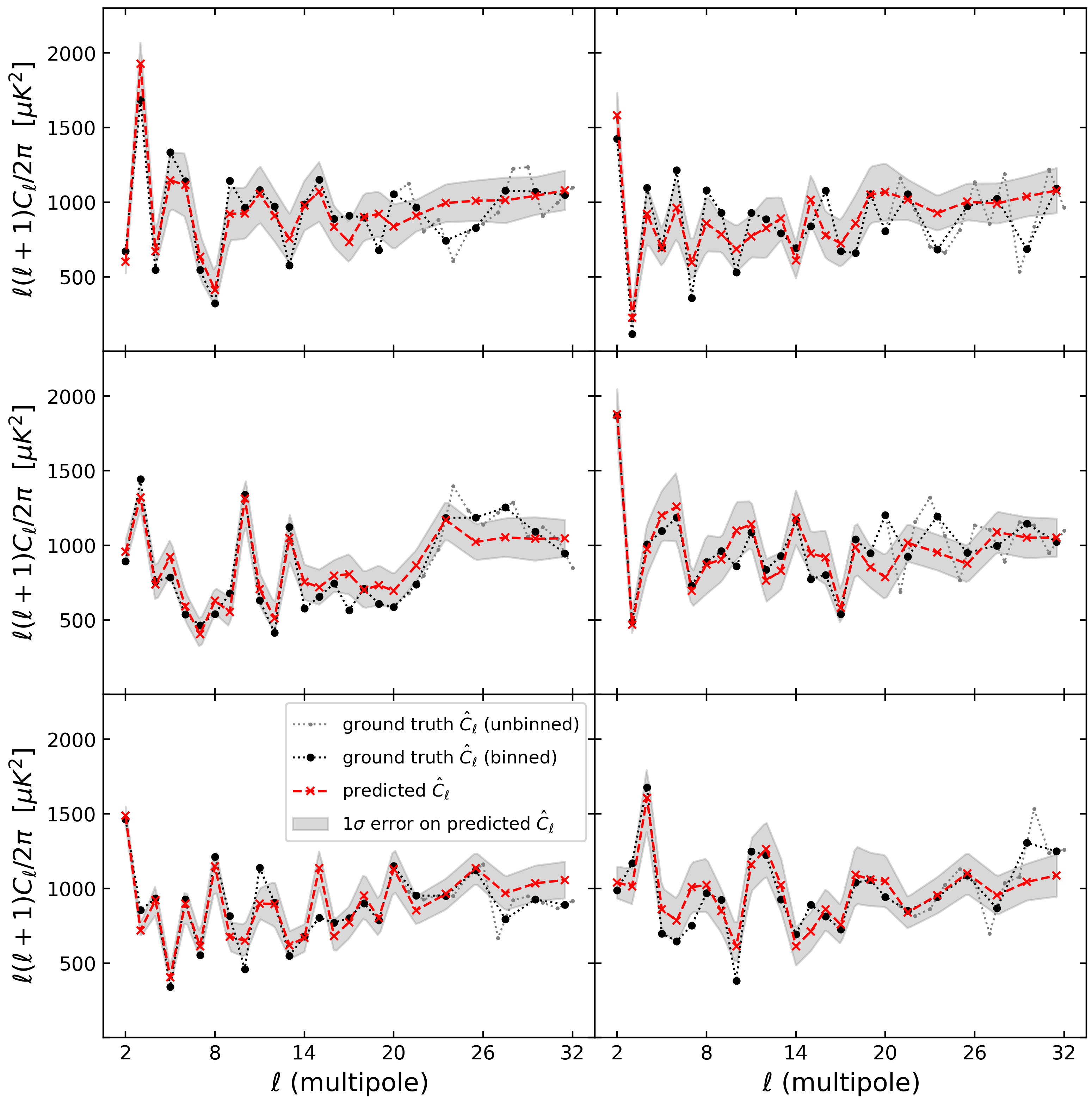}
    \caption{Some examples of full-sky $\hat{C_\ell}$ predictions made by ANN in case \RN{1}, i.e., for the analysis with the Kp2 mask.
    The predicted $\hat{C_\ell}$ are binned starting from $\ell=21$ with $\Delta \ell = 2$.
    The predictions are in good agreement with the ground truths at lower multipoles.
    At higher multipoles, the predictions smoothen out compared with the ground truths.}
    \label{fig:kp2_preds}
\end{figure}

We train two distinct neural networks -
1.~For estimating the full-sky $\hat{C_\ell}$ using partial-sky $\hat{\Tilde{C_\ell}}$ obtained with the Kp2 mask, and
2.~For estimating the full-sky $\hat{C_\ell}$ using partial-sky $\hat{\Tilde{C_\ell}}$ obtained using the Planck T-mask.
Henceforth, we refer to the analyses using Kp2 mask and Planck T-mask as cases \RN{1} and \RN{2}, respectively.
For both ANNs, the input features are $\mathbf{x} = \mathbf{\Tilde{c}}$ obtained using the respective masks, and the true output values are $\mathbf{\hat{y}} = \mathbf{c}$, as discussed in Section \ref{sec:methodology}.
Considering the discussion in Section \ref{sec:simulations} and noting that the $\hat{C_\ell}$ at $\ell=0, 1$ provide no useful information and can consequently be discarded, our input feature vector is 48 dimensional, and our output prediction vector and the corresponding error prediction vector are both 31 dimensional.
That is, our input partial-sky spectra have $\ell_{\mathrm{max}}=47$, and we predict full-sky spectra in the multipole range: $2 \leq \ell \leq 32$.

We have used \href{https://www.tensorflow.org/}{\textsc{TensorFlow}}\footnote{https://www.tensorflow.org/} \citep{tensorflow2015-whitepaper} machine-learning framework to set up, train, and analyze the ANNs.
Training a neural network is an iterative procedure.
Applying mean normalization pre-processing on input features of the training set for both cases, we transform them to have a similar spread.
This allows an optimization algorithm to efficiently explore the weight-space.
The mean and the scale used to transform the training set are also used to process the validation and test sets to ensure that prediction set features go through the same pre-processing as training ones.
This helps to accurately estimate the reliability of a network.

It is found that an ANN with two hidden layers, each having 1024 neurons with ReLU activation and Concrete dropout implemented in the layers, performs best for prediction of full-sky $\hat{C_\ell}$ in both cases.
The best hyper-parameters of the model were obtained using grid search.
A length scale of $0.1$ was used for Concrete dropout.
Concrete dropout also ensures that appropriate weight regulariser and dropout regulariser are applied based on the length scale and size of the training dataset \citep{gal2017concrete}.
Fig. \ref{fig:our_ann} shows a schematic of the ANN architecture used in this work.
The training takes about 35 minutes on a personal computer with Intel\textsuperscript{\textregistered} Core\textsuperscript{\texttrademark} i5-6200U (@ 2.30GHz$\times$4, 8 GB RAM).
We use \textit{ADAM} as our optimization algorithm with its default parameters and also use mini-batches while training the ANNs.
The performance of ANNs on the corresponding validation sets are also tracked.

After training the networks, we obtain predictions on the test sets.
By iterating the network 200 times for the same set of input features, we predict the uncertainty in our $\hat{C_\ell}$ estimates utilizing the procedure described in section \ref{sec:methodology}.
We also bin the predicted full-sky spectra $\ell=21$ onward with a bin width of two in order to get an unbiased estimate at larger multipoles.
Thus, six bins are obtained between $21 \leq \ell \leq 32$.
We assign the binned $\hat{C_\ell}$ to the central multipole of the corresponding bin.

\section{Results} \label{sec:results}
\begin{figure}
    \centering
    \includegraphics[width=.48\textwidth]{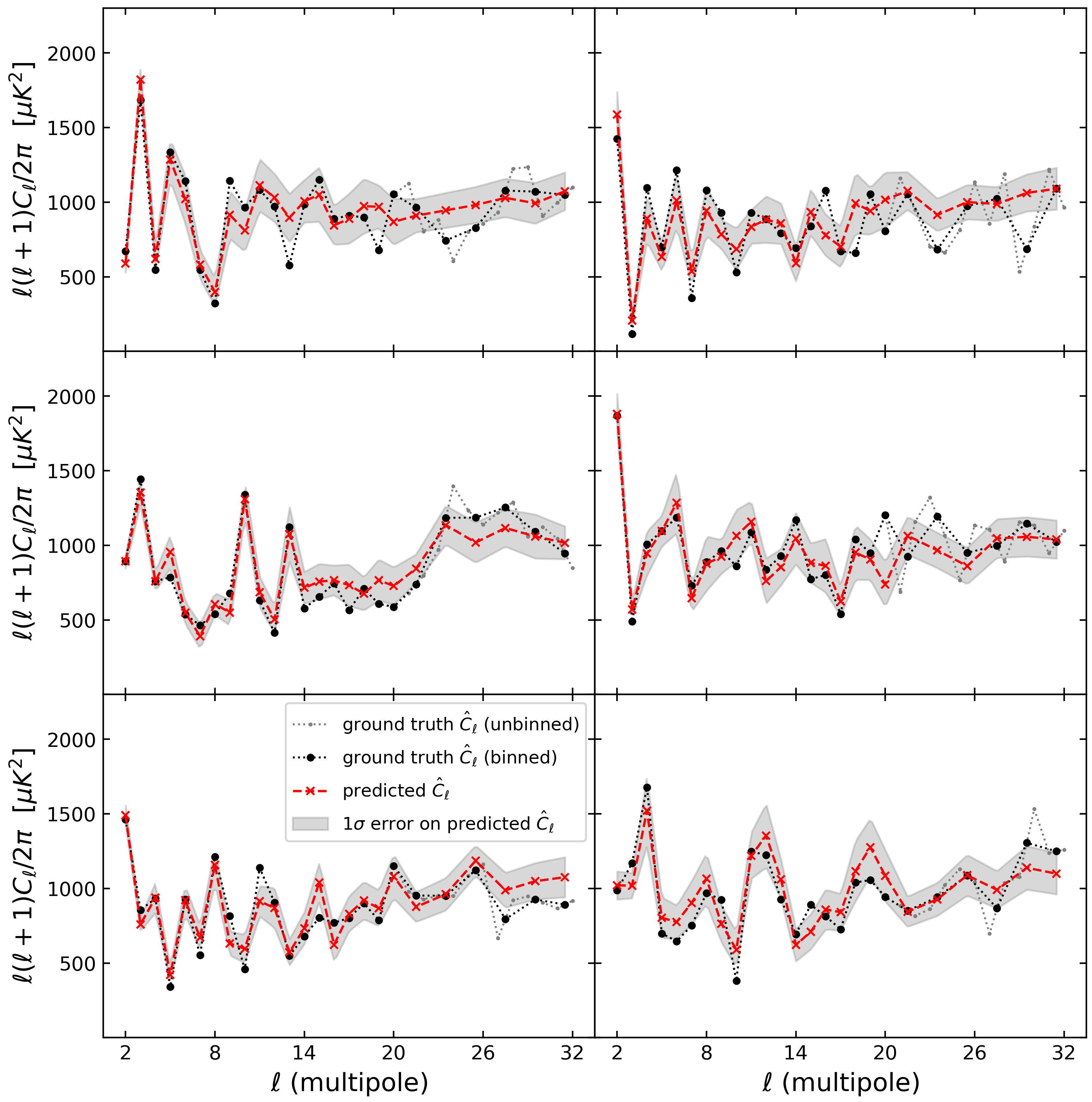}
    \caption{Same as Figure \ref{fig:kp2_preds} but for case \RN{2}, i.e., the analysis with the Planck T-mask.}
    \label{fig:plk_preds}
\end{figure}

In this section, we discuss the results from simulations depicting the use of two different masks.
We mask the simulated CMB maps once with Kp2 mask and once with Planck T-mask, get the required power spectra, and create corresponding training, validation, and test sets using the procedure described in Section \ref{sec:simulations}.
The trained ANNs predict full-sky CMB angular power spectra on the corresponding test sets, followed by the implementation of binning strategy outlined in Section \ref{sec:training} to get unbiased predictions.

Fig. \ref{fig:kp2_preds} and Fig. \ref{fig:plk_preds} show some predictions of the full-sky angular power spectrum made by the ANNs on the test sets along with the ground truth (original) power spectrum for cases \RN{1} and \RN{2}, respectively.
To ensure a fair comparison, the example predictions in both cases are chosen in a way that the ground truths remain same in respective panels of the corresponding figures.
The $1 \sigma$ error estimated by the network is also shown.
At lower multipoles, it is clearly evident that the $\hat{C_\ell}$ predictions made by neural networks in both cases almost trace the ground truth $\hat{C_\ell}$.
This implies that the reconstruction is considerably accurate at those multipoles, which is a result of the ANNs being able to learn a good mapping.
However, the mask leads to a greater loss of information at higher multipoles (refer Fig. \ref{fig:procedure_mask}).
A clear sign of this is indicated by higher aleatoric uncertainty at larger multipoles in our predictions.
Consequently, we observe that the ANNs are unable reconstruct the features or fluctuations in power spectra efficiently at high multipoles.

\begin{figure}
    \centering
    \includegraphics[width=.48\textwidth]{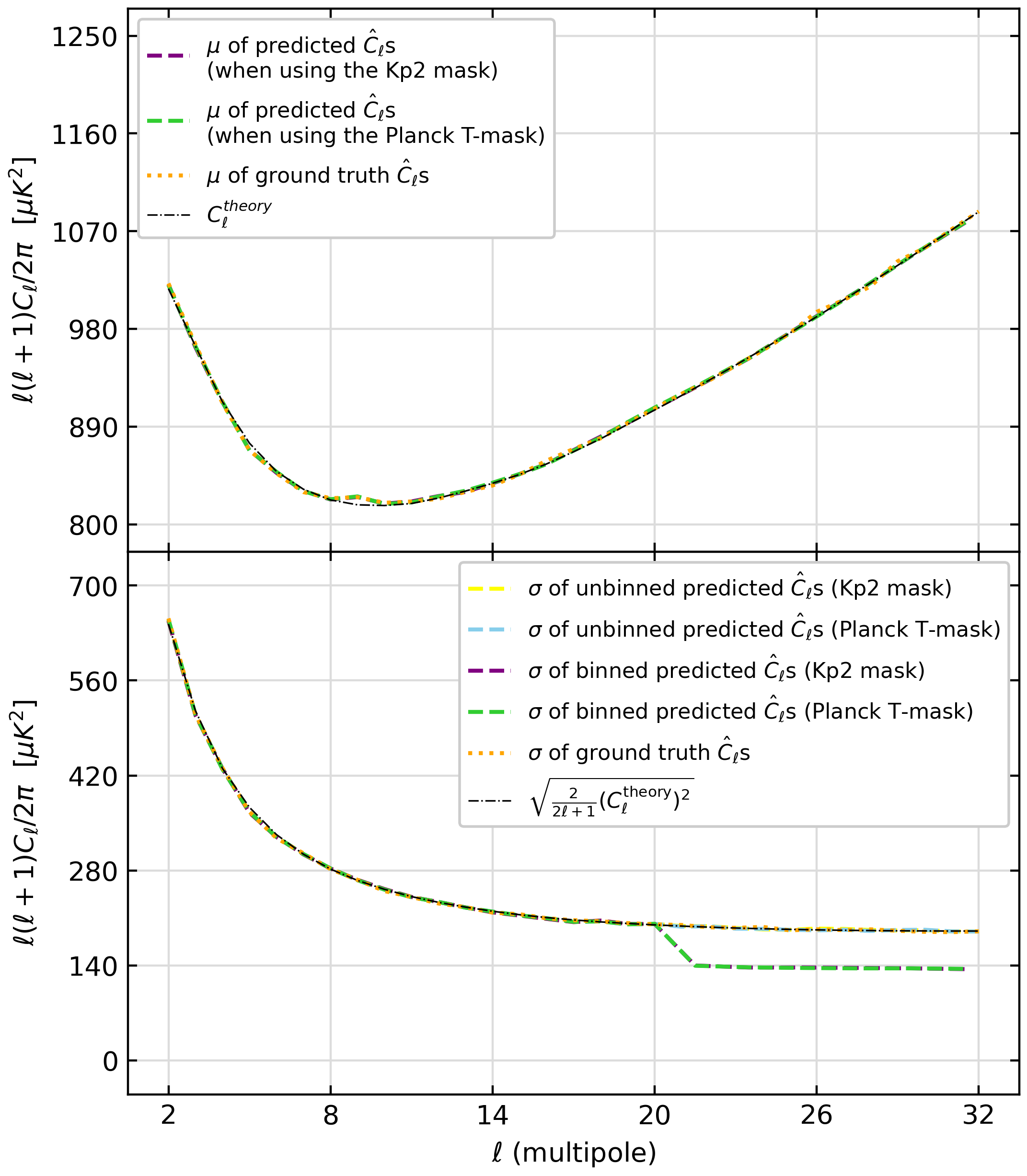}
    \caption{Figure showing the statistics - mean (top) and standard deviation (bottom) - of the predicted full-sky $\hat{C_\ell}$s and those of the ground truths for the analyses using both masks.
    The theoretically expected statistics are also shown.
    Both mean and standard deviation of the predictions are in good agreement with $C_\ell^{\mathrm{th}}$ and cosmic variance respectively.
    The standard deviation of the binned predictions is slightly smaller than the square root of the cosmic variance $\ell=21$ onwards due to binning of the predictions at higher multipoles.
    The standard deviation of unbinned predictions are also plotted for reference; they trace the cosmic variance.}
    \label{fig:meanrecon}
\end{figure}

\begin{figure}
    \centering
    \includegraphics[width=.48\textwidth]{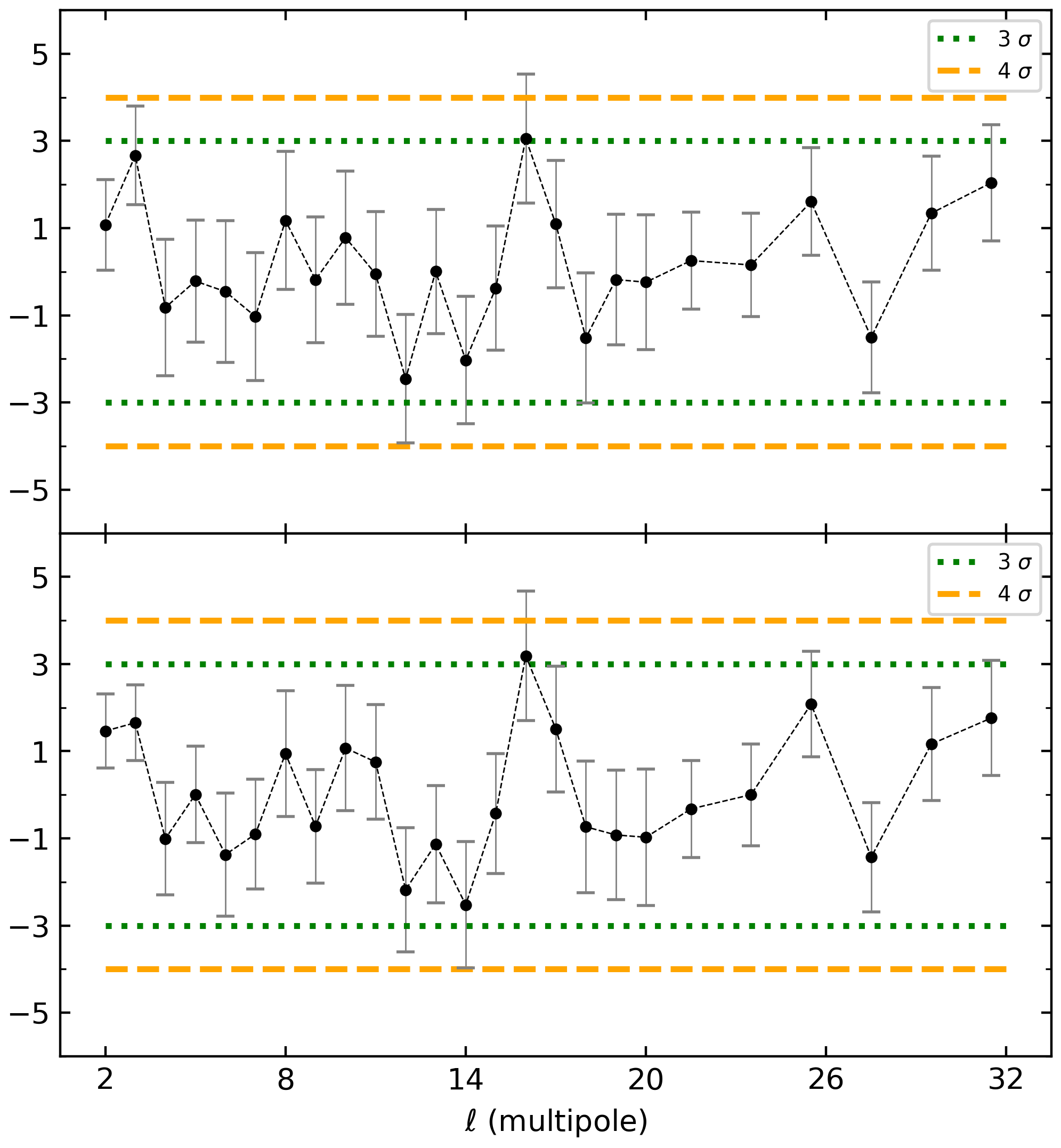}
    \caption{Figure showing mean and SEM of the $\hat{C_\ell}$ differences ($\mathbf{c}^{\textrm{ground-truth}}$ - $\mathbf{c}^{\textrm{predicted}}$) on the test set for case \RN{1}, i.e., when using the Kp2 mask (top), and for case \RN{2}, i.e., when using the Planck T-mask (bottom).
    The mean is less than $3 \sigma$ at all of the multipoles except at $\ell=16$ when using both masks, where it is slightly higher than $3 \sigma$.}
    \label{fig:sem}
\end{figure}

\begin{figure}
    \centering
    \includegraphics[width=.48\textwidth]{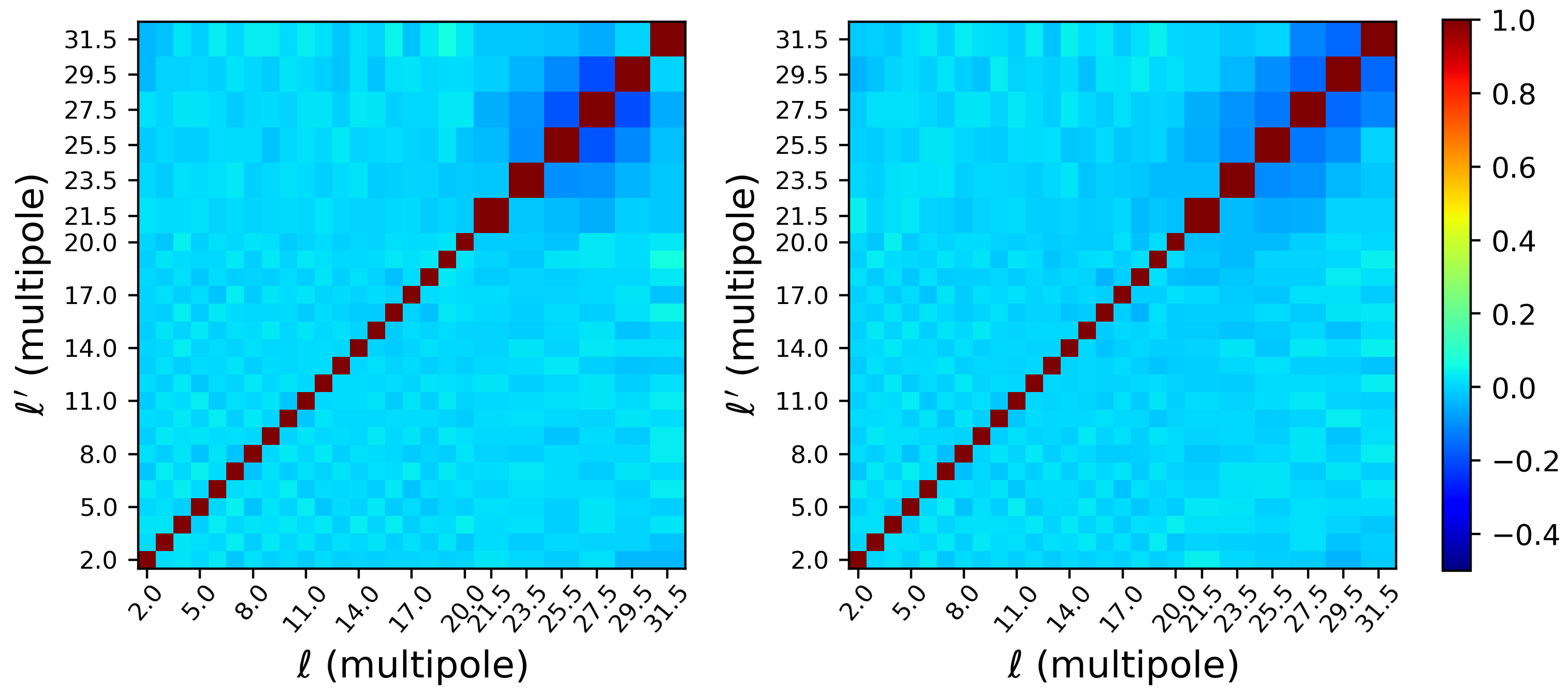}
    \caption{Correlation matrices of the predicted full-sky $\hat{C_\ell}$s for case \RN{1}, i.e., for the analysis with the Kp2 mask (left), and for case \RN{2}, i.e., for the analysis with the Planck T-mask (right).
    The correlations are negligible at lower multipoles.
    At higher multipoles, some correlations are present.
    The magnitude of correlations obtained when using the Planck T-mask are lower than those obtained with the Kp2 mask.}
    \label{fig:corrmat}
\end{figure}

In order to statistically evaluate the performance of predictions made by the ANNs, we compute the mean and standard deviation of the predictions for both cases and compare them with those of the ground truths and theoretical $C_\ell^{\mathrm{th}}$ in Fig. \ref{fig:meanrecon}.
In the top panel, mean of the $10^4$ predicted full-sky $\hat{C_\ell}$s traces the $C_\ell^{\mathrm{th}}$ and mean of the original ground truth $\hat{C_\ell}$s in both cases.
Standard deviation of the unbinned predictions also trace square root of the cosmic variance and standard deviation of the ground truths as shown in the bottom panel.
The small dip in the standard deviation of the binned predictions at higher multipoles is caused by binning of the predicted $\hat{C_\ell}$, $\ell=21$ onwards in both cases.
The ANNs achieve this feat by learning from training examples solely on the basis of cost function, without having any explicit information about $C_\ell^{\mathrm{th}}$ or cosmic variance.

The power spectra differences ($\mathbf{c}^{\textrm{ground-truth}}$ - $\mathbf{c}^{\textrm{predicted}}$) for all examples in the corresponding test sets are obtained.
Fig. \ref{fig:sem} plots the mean of these differences along with standard error of the mean (SEM) for cases \RN{1} and \RN{2} in top and bottom panels, respectively.
In both the cases, mean of the $\hat{C_\ell}$ differences is below $3 \sigma$ at almost all of the multipoles.
At $\ell=16$ for both cases \RN{1} and \RN{2}, mean of the $\hat{C_\ell}$ differences is slightly above $3 \sigma$.
This indicates that mean of the predictions on test set for both cases provide an accurate measure of the true population mean.
In simpler words, it can be inferred that predictions of the full-sky spectra obtained using our neural networks are unbiased.

We further present the multipole-space correlation matrices of the full-sky $\hat{C_\ell}$ predicted by ANNs for cases \RN{1} and \RN{2} in left and right panels of Fig. \ref{fig:corrmat}, respectively.
The correlations are absent at lower multipoles up to $\ell \lesssim 24$ for both cases.
For the case \RN{1} predictions, higher multipoles have a maximum correlation of about $19\%$ in four blocks near $\ell=27.5, 29.5$ and about $11\%$ in two blocks near $\ell=27.5$, while all other regions have correlations less than $10\%$.
For the predictions in case \RN{2}, there exists a correlation of about $16\%$ in four blocks near $\ell=29.5, 31.5$, about $13\%$ in two blocks near $\ell=29.5$, and about $12\%$ in two blocks near $\ell=27.5$ with rest of the regions having correlations less than $10\%$.
We observe that the magnitude of correlations are overall lower in case \RN{2} than case \RN{1}.
This is likely the result of there being approximately $3\%$ lesser masked pixels in the Planck T-mask than Kp2 mask, which leads to somewhat more information in the corresponding $\hat{\Tilde{C_\ell}}$.
The corresponding ANN is able to use this information to make predictions with comparatively lesser correlations.

Finally, covariance matrices in multipole space of the predictions made by the ANNs and the estimations given by the method of unapodised pseudo-$C_\ell$ on the test sets are shown side by side for comparison in Fig. \ref{fig:covmat_kp2} and Fig. \ref{fig:covmat_plk}, respectively for cases \RN{1} and \RN{2}.
Visualization of the matrices using a logarithmic color scale brings out features of the matrices as seen in the figures.
At first glance, one discerns that covariance matrices for the two methods are structured differently from one another.
Covariance matrices for the pseudo-$C_\ell$ estimates have a smooth structure with growing covariances at larger multipoles.
The corresponding matrices for ANN predictions have small covariances in all regions except at lower multipoles and in an area around the diagonal at higher multipoles where we observe moderate-level covariances.
Thus, our method using artificial neural networks also attains significantly lower levels of covariance on the unbiased full-sky $\hat{C_\ell}$ predictions than the pseudo-$C_\ell$ estimates for both cases \RN{1} and \RN{2}.
This coincides with the overall aim of obtaining unbiased and uncorrelated estimates of the full-sky power spectra with as little multipole-space covariance as possible.
Alongside, one observes lower levels of covariance in low-high and high-low multipolar regions for both cases.
This exhibits that our ANN-based technique significantly reduces the low-high multipole coupling compared to pseudo-$C_\ell$.

\section{Discussion and Conclusion} \label{sec:conclusions}
In this paper, we have demonstrated that supervised machine learning with Artificial Neural Networks (ANNs) can be employed to estimate the full-sky CMB angular power spectrum ($\hat{C_\ell}$) effectively from the partial-sky power spectrum ($\hat{\tilde{C_\ell}}$).
We have presented an unbiased estimator of the full-sky $\hat{C_\ell}$ and accurately recovered the theoretical CMB power spectrum ($C_\ell^{\mathrm{th}}$).

We have considered two different masks in our analysis - the Kp2 mask and the Planck T-mask.
We exhibit that an ANN with just two hidden layers can be utilized for the purpose.
The optimum number of layers and neurons in a layer were found by training various neural network architectures until we got the best results in both cases.
We have not used detector noise in our simulations, which is a reasonable assumption for temperature and power spectrum analysis over large angular scales at $N_{\mathrm{side}}=16$.
In our analysis, we have also assumed that the theoretical CMB power spectrum is completely known and have simulated the training and test data accordingly as reported in Section \ref{sec:simulations}.

Both masks conceal comparable regions of the sky.
The predictions made by the ANNs on the corresponding test sets show that the estimations are equally good (Fig. \ref{fig:kp2_preds} and \ref{fig:plk_preds}).
By comparing overall statistics of the predictions with those of the ground truths and $C_\ell^{\mathrm{th}}$, we have shown that our ANN does not output negative or arbitrarily high full-sky power spectra at higher multipoles where the information loss due to cut-sky is significant.
The mean and SEM of the $\hat{C_\ell}$ differences ($\mathbf{c}^{\textrm{original}}$ - $\mathbf{c}^{\textrm{predicted}}$) on our test sets for both cases also show that the full-sky $\hat{C_\ell}$ predictions are unbiased (Fig. \ref{fig:sem}).
This helps us accurately recover the mean $C_\ell^{\mathrm{th}}$ and the cosmic variance (Fig. \ref{fig:meanrecon}), while obtaining a decent unbiased estimate of $\hat{C_\ell}$s even at larger multipoles.

\begin{figure}
    \centering
    \includegraphics[width=.48\textwidth]{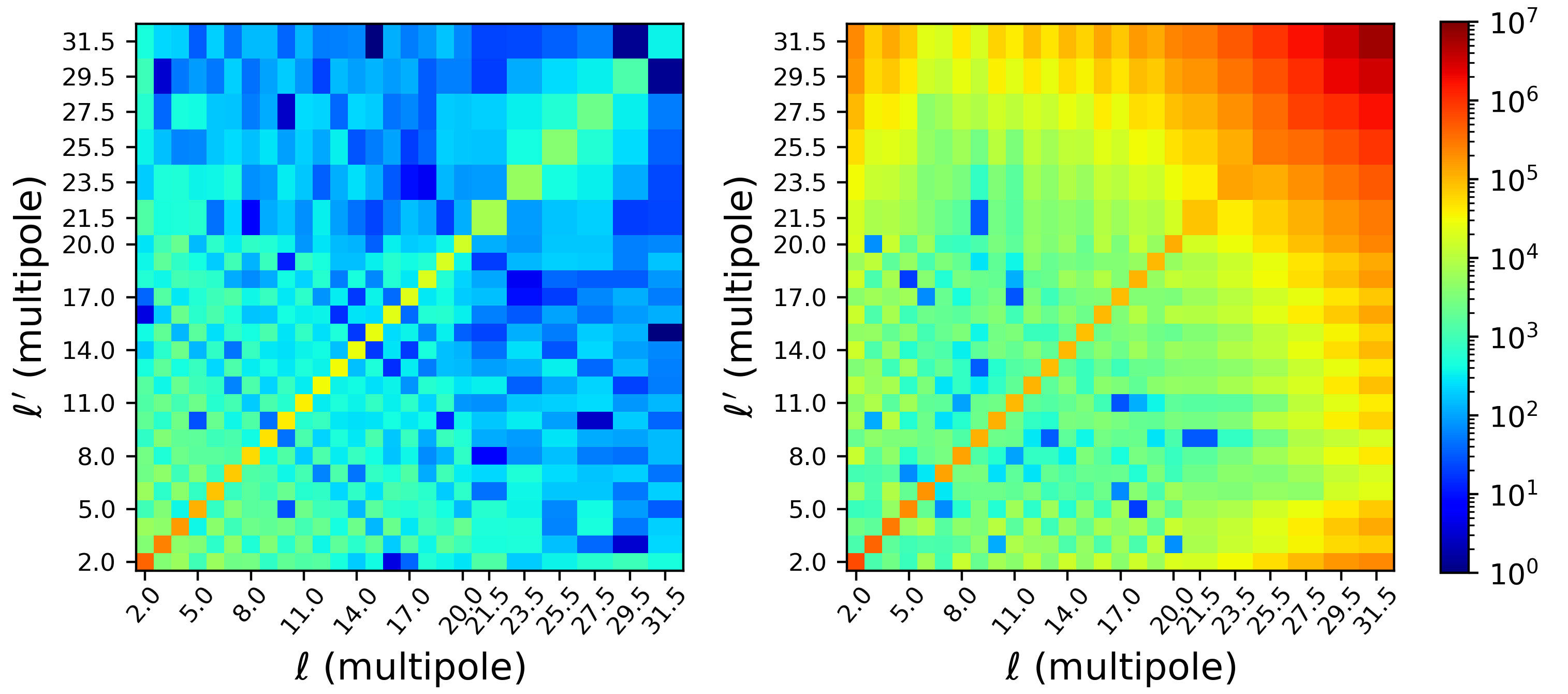}
    \caption{Left panel: Covariance matrix of the predicted $\hat{C_\ell}$s using ANN for case \RN{1}, i.e., when using the Kp2 mask.
    The matrix shows that the predictions have some covariance at lower multipoles and in a region around the diagonal at higher multipoles, while the covariances are much lower in the remaining areas.
    Right panel: Covariance matrix of the estimated $\hat{C_\ell}$s using the pseudo-$C_\ell$ method with the Kp2 mask.
    The matrix has a smooth structure with the covariances being large at higher multipoles.
    Overall, the covariances obtained using the ANN predictions are substantially lower than those observed in pseudo-$C_\ell$ estimates.
    The ANN predictions also have significantly lesser low-high multipole coupling compared to pseudo-$C_\ell$.}
    \label{fig:covmat_kp2}
\end{figure}

\begin{figure}
    \centering
    \includegraphics[width=.48\textwidth]{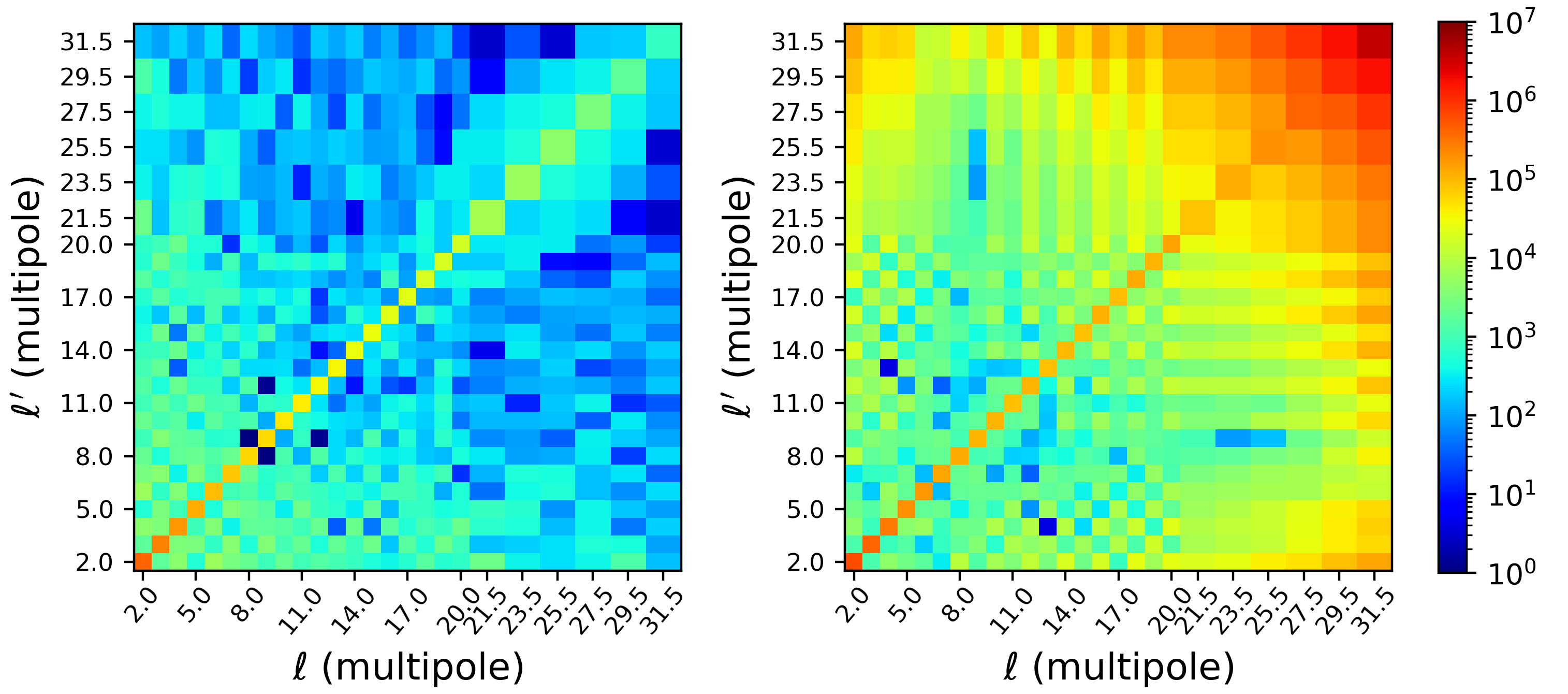}
    \caption{Same as Figure \ref{fig:covmat_kp2} but for case \RN{2}, i.e., when using the Planck T-mask for both methods using ANN and pseudo-$C_\ell$.}
    \label{fig:covmat_plk}
\end{figure}

The correlation matrices of the predicted full-sky spectra in both analyses (Fig. \ref{fig:corrmat}) suggest that the $\hat{C_\ell}$ predictions are mostly uncorrelated with minimal residual correlation at higher multipoles where the partial-sky spectra are extremely correlated owing to the loss of information on masked-sky.
By comparing the covariance matrices of full-sky spectra predictions made by ANNs and those obtained using pseudo-$C_\ell$ method (Fig. \ref{fig:covmat_kp2} and \ref{fig:covmat_plk}), we have shown that ANNs obtain significantly lower covariances at all multipoles.
Additionally, the low-high coupling of the ANN estimates in the multipole space is much lower compared to pseudo-$C_\ell$ estimates.

Following, we summarise the main advantages of our technique:
\begin{enumerate}
    \item A neural network can learn and account for the unique mapping existing between $\hat{\tilde{C_\ell}}$ and $\hat{C_\ell}$ for each particular CMB sky realization.
    It can even learn to compensate for some of the lost information that is generalised across all examples.
    \item We obtain noticeably lesser uncertainties in our $\hat{C_\ell}$ predictions at all multipoles compared to pseudo-$C_\ell$ estimates where the uncertainty is majorly dependent on the sample variance.
    Moreover, the estimates are very precise at lower multipoles.
    This is achieved while also accurately retaining the theoretically expected statistics on the ensemble of estimates for multifarious sky realizations.
    \item The covariance matrix of our predictions show significantly lesser levels of covariance compared to pseudo-$C_\ell$.
    The low-high coupling is also substantially minimized.
\end{enumerate}

\citet{gruetjen2017using} show that inpainting, using a neighbouring-pixel-averaging method, applied to small-sized excluded regions and at the boundary regions of the galactic cut also lead to power spectra estimates that outperform pseudo-$C_\ell$ estimates.
A detailed comparison of our method with the above will be done in a future study.
Deep-learning based approaches currently in emergence also aim to recover the CMB angular power spectrum.
\citet{Petroff_2020} use a spherical CNN for foreground cleaning and achieve accurate power spectrum reconstruction up to $\ell \sim 900$.
They use a customised extension of the U-Net architecture to the sphere resulting in a large network which requires about 9 days to train on a GPU.
In comparison, our approach requires about 35 minutes to train on a CPU.
\citet{Montefalcone_2021} inpaint the CMB maps using PCNN by cutting out images of $128 \times 128$ pixels from a flat sky projection resulting in a side of approximately 14.65 degrees for each cut-out.
The masks used by them cover about 10\% of these cut-outs.
The usage of small sky-fractions also leads to significant contribution of the sample variance causing the error-bars to be much larger than the cosmic variance at multipoles comparable to the regime of our work.
Thus, our method, demonstrated to work efficiently with larger masks, will lead to better constraints on the cosmological parameters.

In conclusion, our investigation has produced encouraging results and exemplifies the capacity of Artificial Neural Networks as a new alternative technique for estimation of the full-sky CMB angular power spectrum.
A comprehensive analysis for application of our method on high resolution data which includes instrumental noise will be the subject of a subsequent study of this ANN technique.
A useful advantage of our approach is that it does not require a machine with very high computational power.
Even for a higher resolution analysis, ANNs will be reliably fast in convergence as they can learn complex functional mappings between high dimensional input and output features by using just a few layers in the network with much-advanced optimization algorithms.
In future work, we will also explore the effect of unknown theoretical CMB power spectrum.

\section*{Acknowledgements}
The authors would like to thank Vipin Sudevan for his help.
The authors also thank the anonymous referee for insightful suggestions that helped us to improve the manuscript.
P.C. would like to thank Nirnay Roy and Prashant Shukla for helpful discussions at various stages of this work.
The authors would like to acknowledge the use of open-source packages \href{https://healpix.sourceforge.io/}{\textsc{HEALPix}}\footnote{https://healpix.sourceforge.io/} and \href{https://www.tensorflow.org/}{\textsc{TensorFlow}}\footnote{https://www.tensorflow.org/}, and thank the respective groups.

\section*{Data Availability}

The authors will readily share the data underlying this work on receiving a reasonable request.



\bibliographystyle{mnras}
\bibliography{ref} 





\bsp	
\label{lastpage}
\end{document}